\documentclass{amsart}
\usepackage[T1]{fontenc}
\usepackage{amsfonts}
\usepackage{mathrsfs}
\usepackage{mathtools}
\usepackage{subcaption}
\usepackage{amssymb}
\usepackage[colorlinks=true,linkcolor=blue,allcolors=blue
]{hyperref}
\usepackage[noabbrev,capitalize,nameinlink]{cleveref}
\usepackage{amsmath,blkarray,booktabs,bigstrut}
\usepackage{cite}
\usepackage{graphicx}
\usepackage{dutchcal}
\usepackage{tikz}
\usepackage{tikz-cd}
\usepackage{pdflscape}
\usepackage{hyperref}
\usepackage{multirow} 
\usepackage{longtable} 
\usepackage{array} 
\usepackage{makecell} 
\usepackage{threeparttable}
\usepackage{booktabs}
\usepackage{comment}
\usepackage{tablefootnote}
\usepackage[utf8]{inputenc}
\usepackage[T1]{fontenc}
\usepackage{booktabs}
\usepackage[justification=centering]{caption}
\usepackage{array, caption, threeparttable}
\usepackage[font=small,labelfont=bf,labelsep=none]{caption}
\numberwithin{equation}{section}

\newcommand{\ud}{\,\mathrm{d}} % 2/6/25

\begin{document}
\include{psfig}
\title[The Beta-Generalized Lindley Distribution for Wind Speed]{Beta-Generalized Lindley Distribution: A Novel Probability  Model for Wind Speed}

\author{Tiantian Yang and Dongwei Chen}
\thanks{Department of Mathematics and Statistical Sciences, University of Idaho, Moscow, ID, USA, 83843; Email: tyang@uidaho.edu}
\thanks{Department of Mathematics, Colorado State University, Fort Collins, CO, US, 80523; Email: dongwei.chen@colostate.edu}
\begin{abstract}
Wind speed distribution has many applications, such as the assessment of wind energy and building design. Applying an appropriate statistical distribution to fit the wind speed data, especially on its heavy right tail, is of great interest. 
In this study, we introduce a novel four-parameter class of generalized Lindley distribution, called the beta-generalized Lindley (BGL) distribution, to fit the wind speed data, which are derived from the annual and long-term measurements of the Flatirons M2 meteorological tower from the years 2010 to 2020 at heights of 10, 20, 50, and 80 meters. 
In terms of the density fit and various goodness-of-fit metrics, the BGL model outperforms its submodels (beta-Lindley, generalized Lindley, and Lindley) and other reference distributions, such as gamma, beta-Weibull, Weibull, beta-exponential, and Log-Normal. 
Furthermore, the BGL distribution is more accurate at modeling the long right tail of wind speed, including the $95^{th}$ and $99^{th}$  percentiles and Anderson–Darling statistics at different heights. Therefore, we conclude that the BGL distribution is a strong alternative model for the wind speed distribution.
\end{abstract}
\thanks{Corresponding author: Dongwei Chen. Email address: dongwei.chen@colostate.edu}
\maketitle

%\tableofcontents

\section{Introduction}
Better knowledge of wind speed distribution is of high interest for many applications. For example, solid wind speed analysis can improve air transport safety since high wind speed may cause turbulence and associated problems at taking off and landing \cite{pobovcikova2017application}. A good estimation of the wind speed distribution can also contribute to risk assessment of insurance \cite{khanduri2003vulnerability},  pollutant dispersion control in urban areas \cite{sim2024distributions}, and agricultural management \cite{dukes2006effect}. Furthermore, it contributes to wind engineering such as building and bridge design \cite{baker2007wind}, since defining the design wind speed is a key step in evaluating the wind loading of structures and their safety under the wind \cite{zhang2018extreme}. In particular, accurate estimation of wind speed distribution is critical to the assessment of wind energy potential, the site selection of wind farms, and the operation management of wind power conversion systems \cite{li2010application, qin2023topological}, since researchers often use the wind speed distribution to estimate the wind energy potential and assess the average annual wind energy yield of a wind turbine at a certain location before its installation \cite{jung20183d}. Therefore, it is necessary to estimate the wind speed distribution accurately. 

The statistical distribution of wind speed has its characteristics:  the right tail is heavy. Many probability distributions have been applied to fit the wind speed data, such as the two-parameter Log-Normal distribution \cite{garcia1998fitting}, two-parameter Weibull distribution \cite{dorvlo2002estimating}, two-parameter generalized Lindley distribution \cite{arslan2017generalized}, three-parameter extended generalized Lindley distribution \cite{kantar2018wind},  upper-truncated Weibull distribution \cite{kantar2015analysis}, new alpha power transformed power Lindley distribution \cite{ahsan2022new}, two-parameter gamma distribution and one-parameter Rayleigh distribution \cite{wang2016wind}. Among the distributions used to fit the wind speed, the top four most used distributions are the two-parameter Weibull (W), two-parameter Log-Normal (LogN), two-parameter gamma (GAM), and one-parameter Rayleigh (R) distributions \cite{jung2019wind}. And by far, the most frequently used statistical distribution is the two-parameter Weibull distribution \cite{jung2019wind}, denoted as $W(\alpha, \lambda)$, where its shape ($\alpha$) and scale ($\lambda$) parameters vary with location and season \cite{sim2024distributions}. Note that the one-parameter Rayleigh distribution is a special case of Weibull distribution when $\alpha$ = 2. Interested readers may refer to \cite{wang2016wind, jung2019wind} for more information about a summary of the probability distributions used for wind speed. In this work, we introduce a novel four-parameter statistical model for wind speed, which outperforms other reference distributions and is known as the beta-generalized Lindley (BGL) distribution \cite{oluyede2015new}. 

The BGL model has four parameters $(\alpha, \lambda, a, b)$ and is often used to fit data from a long-tail distribution. Its probability density function (PDF) is 
\begin{equation*}
\begin{split}
    f_{BGL}(x;\alpha,\lambda, a, b) =& \frac{\alpha \lambda^{2}(1+x)e^{-\lambda x}}{B(a,b)(1+\lambda)} \left[1-\frac{1+\lambda +\lambda x}{1+\lambda}e^{-\lambda x}\right]^{a\alpha -1} \\
    & \times \left\{1-\left[1-\frac{1+\lambda +\lambda x}{1+\lambda}e^{-\lambda x}\right]^{\alpha}\right\}^{b-1},
\end{split}
\end{equation*}
where $x \geq 0$, $\alpha>0$, $\lambda>0$, $a>0$, and $b>0$; refer to Section \ref{section3} in this paper for the details of the BGL density and other statistical properties, or the original BGL paper \cite{oluyede2015new}. The four-parameter BGL distribution is proven to have good statistical properties and can capture different shapes of the density function, especially for the long tail \cite{oluyede2015new}. Hence, it is suitable to fit the wind speed data.

This paper is organized as follows. In Section \ref{section2}, we introduce the background of the used wind speed data and its data preprocessing procedure. Section \ref{section3} introduces the statistical properties of beta-generalized Lindley (BGL) distribution, other reference distributions, and multiple goodness-of-fit evaluation metrics used in this study. In Section \ref{section4}, we compare the performance of BGL distribution with other reference distributions on the long-term wind speed data from the years 2010 to 2020 (2010-2020) at four heights (10, 20, 50, 80 meters) using different evaluation criteria; we also investigate the performance for selected annual wind speed in the years 2010, 2015, 2020 at height 80 meters (m). The results show that the BGL distribution performs better, especially on the right tail, including the $95^{th}$ and $99^{th}$ percentiles. Finally, we conclude our results and propose future work in Section \ref{section5}.

\section{Data Background and Preprocessing} \label{section2}

In this study, we use the wind speed data (m/s) derived from the measurements of the Flatirons M2 meteorological tower \cite{jager1996nrel}, located about 8 km south of Boulder, Colorado, USA. The tower is about 82 meters tall and located at 39.9106° N and 105.2347° W, with its base at 1855 m above the mean sea level. We use hourly averaged wind speed data from 00:00:00 of January 1st, 2010, to 23:00:00 of December 31st, 2020, measured at four heights of the M2 tower, ranging from 10, 20, 50, and 80 m above the ground, respectively. The original wind speed data can be downloaded at \cite{M2data}. We conduct data preprocessing as some timestamps in the original data are repeated, and the corresponding wind speeds are marked at -99999 to show the abnormality. However, there were only fourteen repetitions during the period used, and we deleted these repetitions. The total number of observations for each height for the long-term wind speed data for the years 2010 to 2020 (2010-2020) is 96,432. 

The descriptive statistics of the wind speed data (m/s) at heights 10, 20, 50, and 80 m for the years 2010-2020 are listed in the Panel A of Table \ref{tab:table1_ds}, including the sample size ($n$), minimum (min), maximum (max), median, mean, variance, skewness, and kurtosis.
The sample skewness and kurtosis are calculated by %using the method of moments given by
$$ \text{skewness}  = \frac{\frac{1}{n} \sum_{i=1}^n (x_i - \bar{x})^3}{\left[ \frac{1}{n} \sum_{i=1}^n (x_i - \bar{x})^2 \right]^{3/2}}, \ \text{and}  \ \text{kurtosis}  = \frac{\frac{1}{n} \sum_{i=1}^n (x_i - \bar{x})^4}{\left[ \frac{1}{n} \sum_{i=1}^n (x_i - \bar{x})^2 \right]^{2}}. $$ 
The skewness is zero for data sampled from the normal distribution, while the kurtosis is three. A positive skewness indicates the data is positively skewed (right-skewed). When the kurtosis exceeds three, the distribution will have a higher peak and heavier tail than the normal distribution. In this work, we calculate the skewness and kurtosis using the \textit{moment} package of R.

From Panel A of Table \ref{tab:table1_ds}, we find that at each height for the years 2010-2020, the median wind speed is less than the mean, the mean is significantly smaller than the maximum, the skewness is positive, and the kurtosis is greater than three, which indicates the shape of wind speed at each height is right-tailed and has a high peak. Therefore, the $95^{th}$ and $99^{th}$ percentiles of wind speeds at all heights are also listed in Table \ref{tab:table1_ds} to understand the tails better.

In addition to using the long-term wind speed data, we also evaluate the BGL distribution on the annual wind speed data in the years 2010, 2015, and 2020 at 80 m height, respectively. We choose the 80 m height since wind flows more freely with less friction from obstacles such as buildings and mountains at such height, and many wind turbines are working at 80 m height or above. Panel B of Table \ref{tab:table1_ds} presents the corresponding descriptive statistics for the annual wind speed data. The numbers of observations for the years 2010, 2015, and 2020 are 8,760, 8,760, and 8,784, respectively. We note that similar to the distribution of long-term wind speed, each shape of annual wind speed distribution is right-tailed and has a high peak, since the median is less than the mean, the mean is much smaller than the maximum, the skewness is positive, and the kurtosis is greater than three. 

\section{Materials and Methodology}\label{section3}
In this section, we introduce the beta-generalized Lindley (BGL) distribution, including the cumulative distribution function (CDF), probability density function (PDF),  moments, and maximum likelihood estimation. We then discuss the notations and PDFs of other reference distributions. Finally, we show the goodness-of-fit metrics used in this study. 

\subsection{Beta-Generalized Lindley Distribution} \label{section3.1}
This subsection introduces the four-parameter beta-generalized Lindley (BGL) distribution \cite{oluyede2015new}. The CDF of the beta-generalized distribution class is $$F(x)=\frac{B_{G(x)}(a, b)}{B(a,b)},$$ where 
$B_{G(x)}(a, b)=\int_{0}^{G(x)}t^{a-1}(1-t)^{b-1}\ud t,$
and $\frac{1}{B(a,b)}=\frac{\Gamma(a+b)}{\Gamma(a)\Gamma(b)}$ \cite{eugene2002beta}. The $G(x)$ within $B_{G(x)}(a, b)$ is the CDF of a continuous random variable $X$. When selecting the generalized Lindley (GL) distribution for $G(x)$, we have the CDF of the BGL distribution given by
\begin{equation}
F_{BGL}(x; \alpha, \lambda, a, b)=\frac{1}{B(a,b)}\int_{0}^{F_{GL}(x; \lambda, \alpha)}t^{a-1}(1-t)^{b-1}\ud t, \label{BGL_CDF}
\end{equation}
where $F_{GL}(x;\alpha, \lambda)=\left[1-\frac{1+\lambda +\lambda x}{1+\lambda}e^{-\lambda x}\right]^{\alpha}$ is the CDF of the GL distribution \cite{nadarajah2011generalized}, and its PDF is 
$$f_{GL}(x;\alpha, \lambda)=\frac{\alpha \lambda^{2}}{1+\lambda}(1+x)\left[1-\frac{1+\lambda +\lambda x}{1+\lambda}e^{-\lambda x}\right]^{\alpha-1}e^{-\lambda x}.$$
The GL distribution constitutes a flexible family of distributions in terms of the varieties of shapes, including the Lindley (L) distribution, which is known for its thin tail \cite{lindley1958fiducial}.
By taking the derivative of the CDF as shown in Equation \eqref{BGL_CDF}, we have the PDF of the BGL distribution given by
\begin{equation}\label{clever2}
\begin{split}
     f_{BGL}(x;\alpha,\lambda, a, b) &= \frac{\alpha \lambda^{2}(1+x)e^{-\lambda x}}{B(a,b)(1+\lambda)} \left[1-\frac{1+\lambda +\lambda x}{1+\lambda}e^{-\lambda x}\right]^{a\alpha -1} \\
     &\times \left\{1-\left[1-\frac{1+\lambda +\lambda x}{1+\lambda}e^{-\lambda x}\right]^{\alpha}\right\}^{b-1}, 
\end{split}
\end{equation}
where the input data $x \geq 0$, and the four parameters satisfy $\alpha>0$, $\lambda>0$, $a>0$, $b>0$; $B(a,b)$ is the Beta function given by $\frac{1}{B(a,b)}=\frac{\Gamma(a+b)}{\Gamma(a)\Gamma(b)}$, where $\Gamma(\cdot)$ refers to the gamma function, with $\Gamma(a) = \int_0^\infty t^{a-1}e^{-t} \ud t$ and $\Gamma (m)=(m-1)!$ for a positive integer $m$. For different values for the four parameters $\alpha$, $\lambda$, $a$, and $b$, the BGL density function can demonstrate a variety of shapes, as shown in the BGL distribution paper \cite{oluyede2015new}. 

The $s^{th}$ moment of the BGL distribution with integer $s>0$ is given by $$\mu_s= \int_0^\infty x^s f_{BGL}(x; \alpha, \lambda, a, b) \ud x.$$
There are different versions of series representations for the $s^{th}$ moment of the BGL distribution, depending on whether the three parameters $\alpha$, $a$, and $b$ are integers or non-integers, and $\lambda$ does not affect the sum of moment series.  
For the details of $s^{th}$ moments, please refer to the original BGL distribution paper \cite{oluyede2015new} for more information.

\subsection{Maximum Likelihood Estimation of BGL Distribution}

In this part, we obtain the maximum likelihood estimates (MLEs) of BGL distribution. There are many estimation methods for distribution parameters, such as maximum likelihood estimation and moment estimation \cite{jung2019wind}. Compared to the moment estimation, which only uses information from a few moments, the maximum likelihood estimation uses all information from the data. Furthermore, the maximum likelihood estimation is generally preferred to estimate the shape parameter of Weibull distribution when the sample size is greater than 100 \cite{kantar2015analysis, wang2016wind}. Therefore, we use the maximum likelihood method for parameter estimation in this study. 

Rewrite $1-\frac{1+\lambda +\lambda x}{1+\lambda}e^{-\lambda x}$ from Equation \eqref{clever2} as $V(x)$ for simplicity. Then, the PDF of BGL distribution becomes
    \begin{equation}
f_{BGL}(x; \alpha, \lambda, a, b) = \frac{\alpha \lambda^{2}}{B(a,b)(1+\lambda)}(1+x)e^{-\lambda x}[V(x)]^{a\alpha -1}[1-V^{\alpha}(x)]^{b-1}. 
    \end{equation} 
Let $\{x_1,\cdots,x_n \}$ be a random sample from the BGL distribution. Then, the corresponding likelihood function is 
  \begin{equation*}
       L(\alpha, \lambda, a, b)=\left[\frac{\alpha \lambda^{2}}{B(a,b)(1+\lambda)}\right]^n \prod_{i=1}^{n}(1+x_{i})e^{-\lambda \sum_{i=1}^n x_i} 
       \prod_{i=1}^{n}[V(x_{i})]^{a\alpha -1}
       \prod_{i=1}^{n}\left[1-V^{\alpha}(x_{i})\right]^{b-1},
   \end{equation*}
   
   % \begin{equation*}
   % \begin{split}
   %     L(\alpha, \lambda, a, b)&=\left[\frac{\alpha \lambda^{2}}{B(a,b)(1+\lambda)}\right]^n \times \prod_{i=1}^{n}(1+x_{i})e^{-\lambda \sum_{i=1}^n x_i} \\
   %     & \times \prod_{i=1}^{n}[V(x_{i})]^{a\alpha -1}
   %     \times \prod_{i=1}^{n}\left[1-V^{\alpha}(x_{i})\right]^{b-1},
   % \end{split}
   % \end{equation*}
and the associated log-likelihood function is
   \begin{equation}
   \begin{split}
        \log L(\alpha, \lambda, a, b)
&=n \log\left[\frac{\alpha \lambda^{2}}{B(a,b)(1+\lambda)}\right]+ \sum_{i=1}^{n}\log(1+x_{i})-\lambda \sum_{i=1}^n x_i\nonumber \\
&+(a\alpha -1)\sum_{i=1}^{n}\log V(x_i)+(b-1)\sum_{i=1}^{n}\log[1-V^\alpha(x_i)].
   \end{split}
   \end{equation} 
Hence, the partial derivatives of $\log L(\alpha, \lambda, a, b)$ with respect to each of the four parameters are given by
   \begin{eqnarray*}
\frac{\partial \log L(\alpha, \lambda, a, b)}{\partial a}
&=&n[\Psi(a+b)-\Psi(a)]+\alpha \sum_{i=1}^{n}\log V(x_i),
   \end{eqnarray*}
   \begin{eqnarray*}
\frac{\partial \log L(\alpha, \lambda, a, b)}{\partial b}
&=&n[\Psi(a+b)-\Psi(b)]+\sum_{i=1}^{n}\log[1-V^\alpha(x_i)],
   \end{eqnarray*}
   \begin{eqnarray*}
\frac{\partial \log L(\alpha, \lambda, a, b)}{\partial \alpha}
&=&\frac{n}{\alpha}+a \sum_{i=1}^{n}\log V(x_i)
+(1-b)\sum_{i=1}^{n}\frac{V^{\alpha}(x_i)\log V(x_i)}{1-V^{\alpha}(x_i)},
   \end{eqnarray*}
   and
   \begin{equation*}
   \begin{split}
       \frac{\partial \log L(\alpha, \lambda, a, b)}{\partial \lambda} &=\frac{n(2+\lambda)}{\lambda(1+\lambda)}- \left \{\sum_{i=1}^n x_i +\sum_{i=1}^{n}\frac{\lambda(2+\lambda +x_i+\lambda x_i)x_i e^{-\lambda x_i}}{(1+\lambda)} \right. \\
       &\left. \times \left[\frac{a\alpha -1}{(1+\lambda)V(x_i)}+\frac{\alpha(1-b)[V(x_i)]^{\alpha-1}}{(1+\lambda)(1-V^{\alpha}(x_i))}\right] \right\},
   \end{split}
   \end{equation*}
where $\Psi(a)=\frac{\Gamma^{'}(a)}{\Gamma(a)}$ and  $\Gamma^{'}(a)=\int^\infty_0 t^{a-1}(\log t) e^{-t}\ud t$  is the derivative of gamma function $\Gamma(a)$. The maximum likelihood estimates of parameters $\alpha, \lambda, a, b$ are obtained by maximizing the log-likelihood function and letting the above partial derivatives be zeros. As the nonlinear system $(\frac{\partial \log L}{\partial \alpha},\frac{\partial \log L}{\partial \lambda},\frac{\partial \log L}{\partial a},\frac{\partial \log L}{\partial b})^{T}={\mathbf{0}}$ does not admit any explicit solution, the optimal parameters of maximum likelihood estimates, denoted by $\hat{{\mathbf{\Theta}}} = (\hat{\alpha},\hat{\lambda},\hat{a},\hat{b})$, can be obtained using numerical procedures like Newton-Raphson \cite{kelle2003solving}. The MLEs of parameters for each distribution are computed by maximizing the corresponding log-likelihood function via the R package $\textit{bbmle}$. 

\subsection{Reference Distributions and Evaluation Metrics}
The wind speed distribution has a heavy right tail, and many probability distributions have been applied to model such distribution; interested readers can refer to the wind speed distribution review \cite{jung2019wind} for more details. In this study, we employ a novel four-parameter beta-generalized Lindley (BGL) distribution \cite{oluyede2015new} to fit the wind speed data and compare it with its submodels, including beta-Lindley (BL), generalized Lindley (GL) \cite{nadarajah2011generalized}, and Lindley (L) \cite{lindley1958fiducial} distributions, as well as some other reference distributions, including gamma (GAM), beta-Weibull (BW) \cite{lee2007beta}, Weibull (W), beta-exponential (BE) \cite{nadarajah2006beta}, and Log-Normal (LogN) distributions. 

The four-parameter BGL distribution, denoted as $BGL(\alpha, \lambda, a, b)$, has several special cases, including BL, GL, and L distributions \cite{oluyede2015new}: (1) when $\alpha=1$, the BGL distribution reduces to the three-parameter BL distribution, denoted as $BL(\lambda, a, b)$; (2) when $a=b=1$, the BGL reduces to the two-parameter GL distribution, denoted as $GL(\alpha, \lambda)$; (3) when $\alpha=a=b=1$, the BGL reduces to the one-parameter L distribution, denoted as $L(\lambda)$. On the other side, the BE and W distributions are special cases of the four-parameter BW distribution: (1) when $\alpha=1$, the BW distribution reduces to the three-parameter BE distribution, denoted as $BE(\lambda, a, b)$; (2) when $a=b=1$, the BW distribution reduces to the two-parameter W distribution, denoted as $W(\alpha, \lambda)$. For the two-parameter LogN distribution, denoted as $LogN(\alpha,\lambda)$, the $\alpha$ and $\lambda$ refer to mean and standard deviation, respectively. The two-parameter GAM distribution is denoted as $GAM(\alpha, \lambda)$. The details of these nine distributions, including notation, number of parameters ($p$), and probability density function (PDF), are listed in Table \ref{tab:table1_pdf}. 

The goodness-of-fit metrics used to evaluate density fit in this study include -2log-likelihood statistic, $-2\ln(L)$; Akaike Information Criterion, $AIC=2p-2\ln(L)$; and Bayesian Information Criterion, $BIC=p\ln(n)-2\ln(L)$,  
%and  Consistent Akaike Information Criterion, $AICC=AIC+2\frac{p(p+1)}{n-p-1}$, 
where $L=L(\hat{{\mathbf{\Theta}}})$ is the value of the likelihood function evaluated at the optimal estimates of parameters, $n$ is the number of observations, and $p$ is the number of estimated parameters. CDF-based statistics are also popular goodness-of-fit metrics. We use the Kolmogorov-Smirnov (KS) and Anderson–Darling (AD) statistics to quantify the similarities between the empirical CDF of wind speed data (denoted as $F_n(x)$) and the theoretical CDF of reference distributions being tested (denoted as $F(x)$). Both KS statistics and AD statistics can be considered as distances. The KS statistic quantifies the maximum absolute difference between $F_n(x)$ and $F(x)$. The calculation formula is $$KS = \sup_{x \geq 0} \left|F_n(x) - F(x)\right|.$$ %where $n$ is the sample size.
The AD statistic emphasizes the difference in tails of distributions by assigning larger weights for the tails. Therefore,  AD statistics are more sensitive to capturing the deviations in the tails of distributions compared with other evaluation metrics. With the weight function $w(x)$ be $\frac{1}{F(x)(1-F(x))}$, the AD statistic for wind speed is calculated by 
$$AD = n \int_{0}^{\infty} \left( F_n(x) - F(x) \right)^2 w(x) \ud F(x).$$
Letting $x_{(1)}, x_{(2)}, \dots, x_{(n)} \text{ be ordered data points with } x_{(1)} < x_{(2)} < \dots < x_{(n)}$,  we also have 
$$AD = -n - \sum_{i=1}^{n} \frac{2i-1}{n} \left[ \ln(F(x_{(i)})) + \ln(1 - F(x_{(n+1-i)})) \right].$$ 
The smaller the above evaluation metrics are, the better the density fit is. 

\section{Results and Discussion}\label{section4}
This section uses the BGL distribution to model the wind speed and compares it with its sub-models and other popular reference distributions, as shown in Table \ref{tab:table1_pdf}. 
We evaluate the performance of each distribution from the following four perspectives. 
\begin{itemize}
    \item[$(1)$] The goodness-of-fit test on the long-term wind speed data from the years 2010 to 2020. 
    \item[$(2)$] The modeling performance at different heights (10, 20, 50, and 80 m) from the years 2010 to 2020.
    \item[$(3)$] The modeling accuracy at the distribution right tail from the years 2010 to 2020, including the 95th and 99th percentiles, and AD statistics. 
    \item[$(4)$] The goodness-of-fit test on the annual wind speed data in the years 2010, 2015, and 2020 at 80 m height.
\end{itemize}

\begin{figure}[ht]
\centering
\includegraphics[scale=0.7]{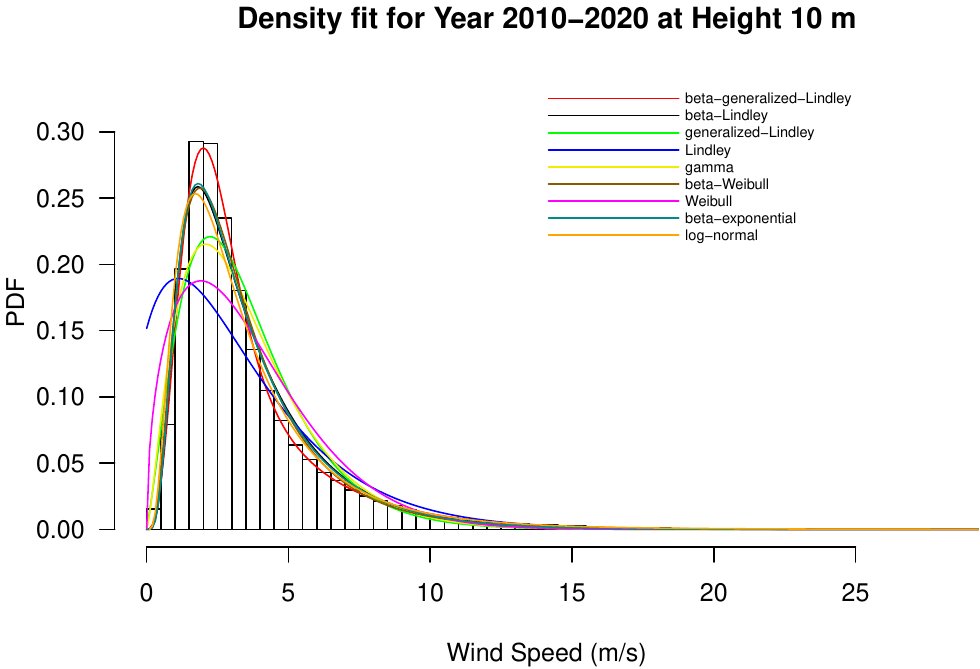}
\caption{. Density fits of nine distributions for the wind speed in the years 2010-2020 at 10 m height}
\label{fig:label10-20_H10}
\end{figure}

\begin{figure}[ht]
\centering
\includegraphics[scale=0.7]{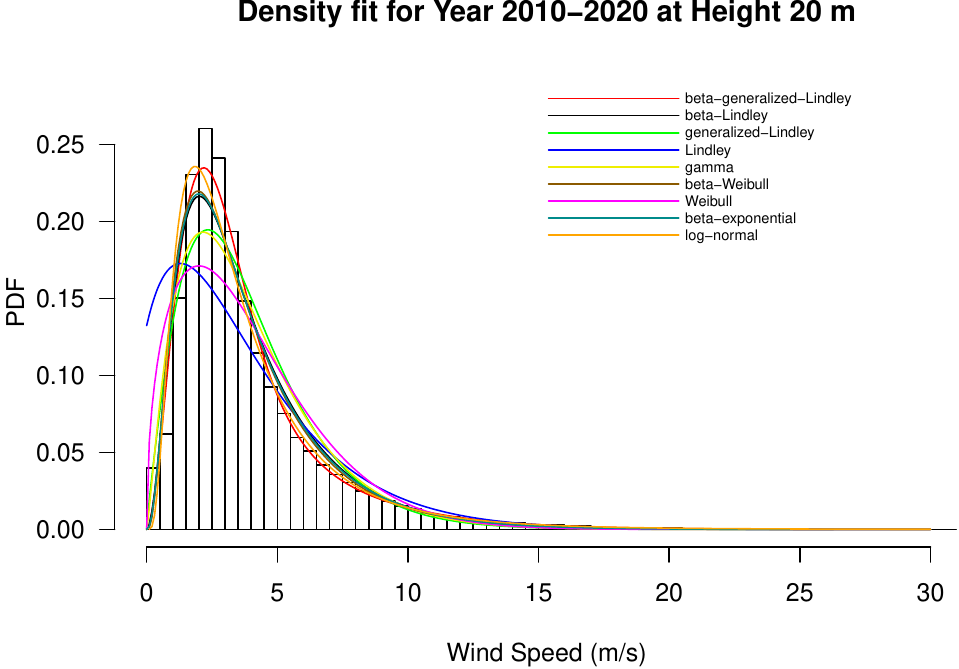}
\caption{. Density fits of nine distributions for the wind speed in the years 2010-2020 at 20 m height}
\label{fig:label10-20_H20}
\end{figure}

\begin{figure}[ht]
\centering
\includegraphics[scale=0.7]{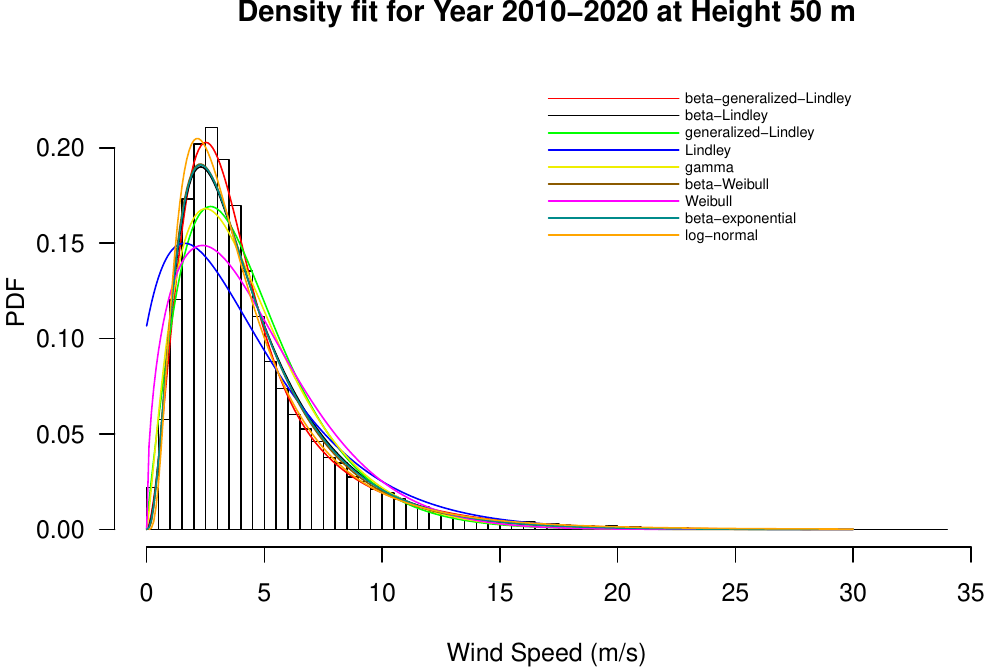}
\caption{. Density fits of nine distributions for the wind speed in the years 2010-2020 at 50 m height}
\label{fig:label10-20_H50}
\end{figure}

\begin{figure}[ht]
\centering
\includegraphics[scale=0.7]{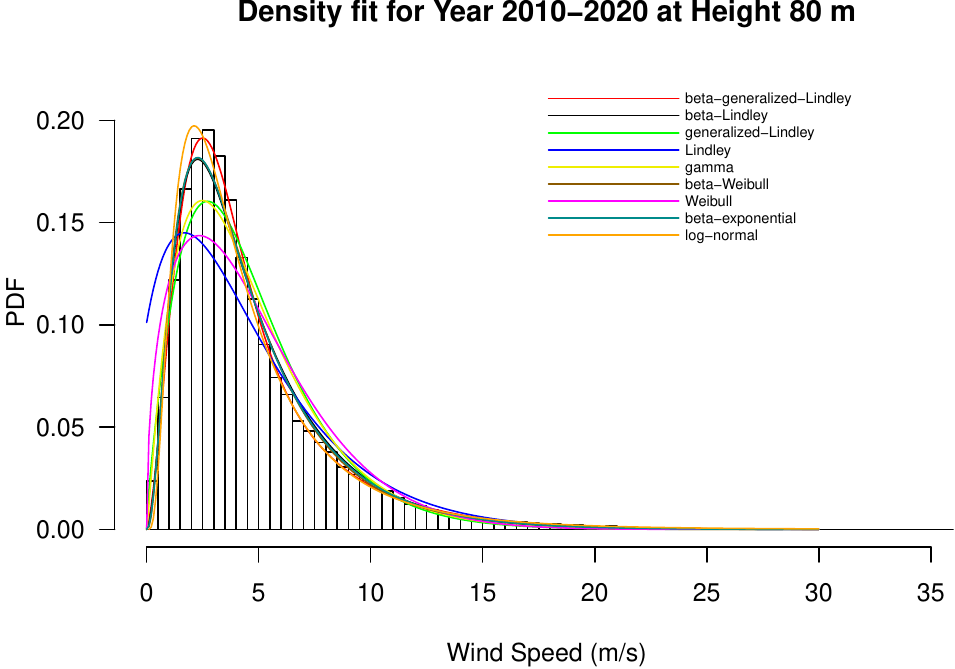}
\caption{. Density fits of nine distributions for the wind speed in the years 2010-2020 at 80 m height}
\label{fig:label10-20_H80}
\end{figure}

We first show the results of long-term wind speed data from the years 2010 to 2020 (2010-2020) at heights 10, 20, 50, and 80 meters (m), including the values of estimated parameters and goodness-of-fit metrics. The results are summarized in Table \ref{tab:table1_H10}, where the distributions are listed in the ascending order of the $-2ln(L)$ value. At the 10 m height, the goodness-of-fit metrics including $-2\ln(L)$, AIC, BIC, KS, and AD for the BGL distribution are 393606.3, 393614.3, 393652.2, 0.008, and 11.4, respectively, which are the smallest among all the distributions listed in Table \ref{tab:table1_pdf}. This indicates that the BGL distribution outperforms other reference models on the wind speed data at 10 m height. The four lowest-performing distributions are GAM, GL, W, and L; their corresponding evaluation metrics are all larger than the counterparts of the BGL distribution. The distribution analyses for wind speed data from the years 2010-2020 at 20, 50, and 80 m heights also show a similar pattern: all the goodness-of-fit metrics for the BGL distribution outperform the counterparts of other reference models, and the four lowest-ranked distributions are GAM, GL, W, and L. 
Furthermore, we plot the histograms for long-term wind speed data at different heights (10, 20, 50, and 80 m), the fitted BGL density function, and other reference distribution density functions in Figures \ref{fig:label10-20_H10}, \ref{fig:label10-20_H20}, \ref{fig:label10-20_H50}, and \ref{fig:label10-20_H80}, respectively.  It can be seen from Figures \ref{fig:label10-20_H10}- \ref{fig:label10-20_H80} that the BGL exhibits a better fit on the wind speed data than other reference distributions. In particular, the BGL densities capture the high peaks and the long right tails at all heights well. From the above analysis, we conclude that the BGL distribution is more suitable to model the long-term wind speed data at different heights.

To investigate the performance of all distributions in the long right tail, we particularly calculate the estimated $95^{th}$ and $99^{th}$ percentiles from fitted distributions, the observed percentiles of wind speed data from the years 2010-2020 at all heights, and their differences (biases). Table \ref{tab:table1_percentile} lists the results, and the distributions are ordered in ascending by the absolute value of bias for the $95^{th}$ percentile (or 0.95 quantile score). Positive bias indicates overestimation, while negative bias shows underestimation. At the 10 m height, the observed $95^{th}$ and $99^{th}$ percentiles of wind speeds are 8.86 m/s and 13.95 m/s, respectively. The best estimation for the $95^{th}$ percentile is the BGL distribution, with a relatively small overestimated bias of 0.07 m/s. In comparison, the magnitudes of the $95^{th}$ percentile biases of other reference distributions are all greater than 0.38 m/s, significantly larger than the counterpart of the BGL distribution. For the $99^{th}$ percentile estimation at 10 m height, the BGL distribution also has the best performance: it has the smallest absolute value of bias at 0.35 m/s, while the magnitudes of biases of other distributions are all above 0.69 m/s. 
Similar analyses at heights 20, 50, and 80 m also show that the BGL distribution has better estimation on the $95^{th}$ and $99^{th}$ percentiles than other reference distributions, which verifies that the BGL distribution is more accurate at modeling the distribution right tail of wind speed at different heights. Furthermore, as mentioned in Section \ref{section3}, the Anderson–Darling (AD) statistic is more sensitive to the deviations in the distribution tails, and the smaller the AD statistic admits, the better the density fit is. From Table \ref{tab:table1_H10}, one can see that the AD values of BGL distribution at heights 10, 20, 50, and 80 m, ranging from 11.4, 73.9, 13.6, and 7.6, are much smaller than the associated ADs of any other distributions. 
This further verifies that compared to other widely accepted models, the BGL distribution performs better at modeling the right tail of wind speed distribution at different heights. 

\begin{figure}[ht]
\centering
\includegraphics[scale=0.7]{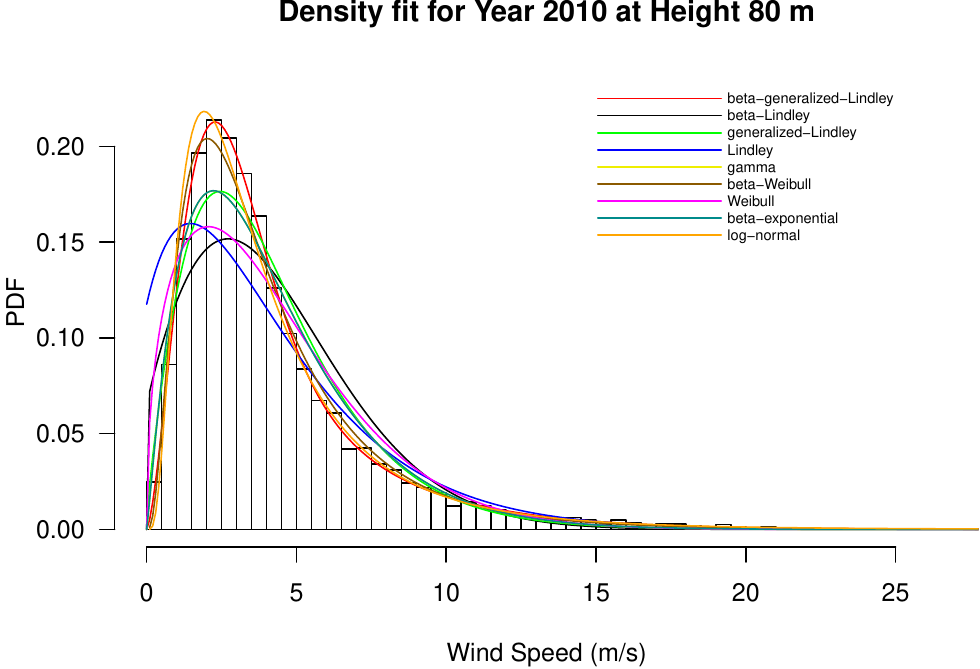}
\caption{. Density fits of nine distributions for the wind speed in the year 2010 at 80 m height}
\label{fig:label10_H80}
\end{figure}

\begin{figure}[ht]
\centering
\includegraphics[scale=0.7]{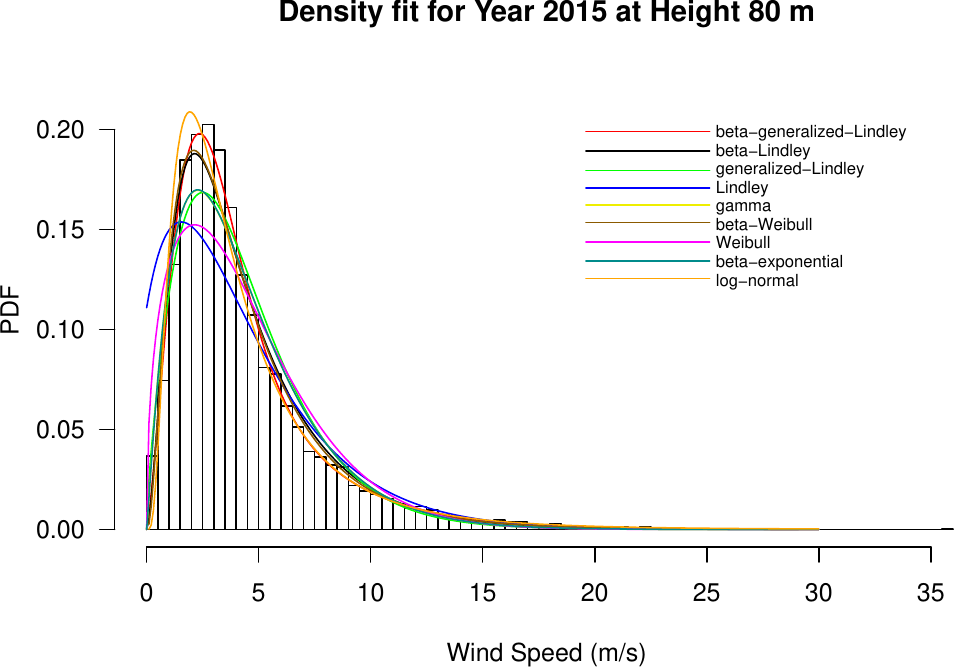}
\caption{. Density fits of nine distributions for the wind speed in the year 2015 at 80 m height}
\label{fig:label15_H80}
\end{figure}

\begin{figure}[ht]
\centering
\includegraphics[scale=0.7]{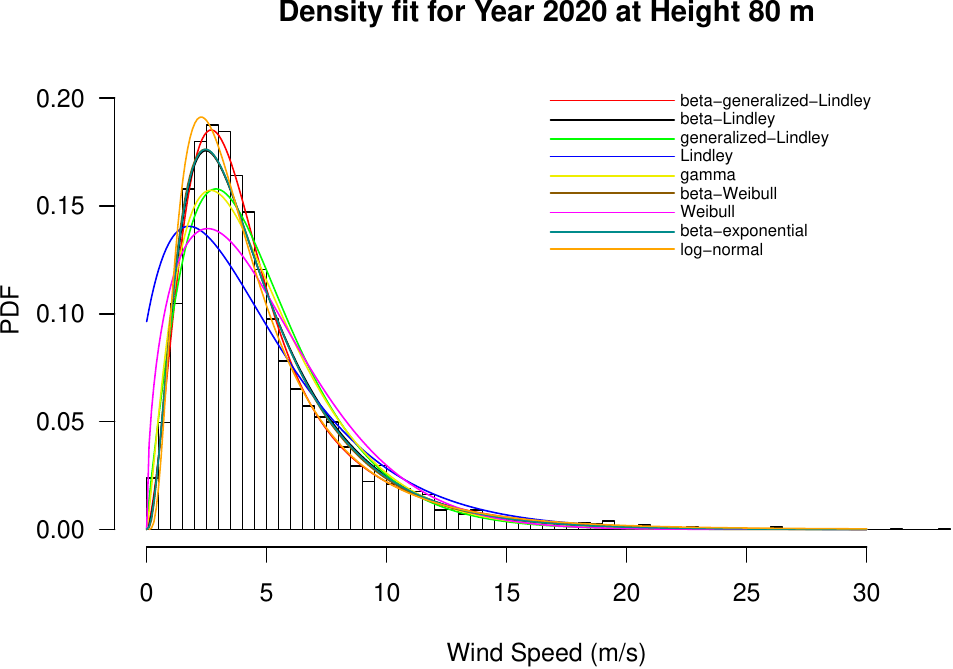}
\caption{. Density fits of nine distributions for the wind speed in the year 2020 at 80 m height}
\label{fig:label20_H80}
\end{figure}

We finish this section by studying the performance of BGL distribution on the annual wind speed data in the years 2010, 2015, and 2020 at 80 m height. The values of estimated parameters and goodness-of-fit metrics are listed in Table \ref{tab:table_H80}. It is noted that compared to the reference distributions, the BGL distribution has the lowest values in the goodness-of-fit metrics in the years 2010, 2015, and 2020, including $-2ln(L)$, AIC, BIC, KS, and AD. This indicates that the BGL distribution is more suitable for modeling the annual wind speed data, which aligns with the best performance of BGL distribution on the long-term wind speed data from the years 2010 to 2020. Moreover, the fitted BGL density function and other reference density functions for the years 2010, 2015, and 2020 at 80 m height are plotted in Figure \ref{fig:label10_H80}, \ref{fig:label15_H80} and \ref{fig:label20_H80}, respectively. It further verifies from Figure \ref{fig:label10_H80}-\ref{fig:label20_H80} that the BGL distribution performs better on the annual wind speed data than other distributions.

In summary, the BGL distribution performs better than other reference distributions on the long-term and annual wind speed data at all different heights. In particular, the BGL distribution is more accurate when modeling the right tail of wind speed distribution.

\section{Conclusion and Future Work}\label{section5}
This study uses the beta-generalized Lindley (BGL) 
distribution to fit the wind speed data with a heavy right tail. For the long-term wind speed data from the years 2010 to 2020 (2010-2020),  the goodness-of-fit metrics, including $-2ln(L)$, AIC, BIC, KS statistic, and AD statistic, show that the BGL distribution outperforms its submodels (beta-Lindley, generalized Lindley, Lindley) and other reference distributions (gamma, beta-Weibull, Weibull, beta-exponential, and Log-Normal) at four different heights (10, 20, 50, and 80 m). Furthermore, the BGL distribution has the smallest values in the Anderson-Darling (AD) statistic and fewer biases in estimating the $95^{th}$ and $99^{th}$ percentiles, which shows that the BGL distribution is more accurate at modeling the right tail of wind speed data. The density fit and goodness-of-fit evaluation in the years 2010, 2015, and 2020 at 80 m height further confirm that the BGL model can also perfectly fit the annual wind speed data compared to other reference distributions. 
Therefore, we conclude that the BGL distribution is a strong alternative model for wind speed distribution estimation and can be applied to many wind-related scenarios. In the future, we will use the BGL distribution to evaluate the numerical weather models on wind speed and revise the output of wind simulations.

\begin{landscape}
    \begin{table*}[!hbt]
  \centering
  \caption{Descriptive statistics of long-term wind speed (m/s) in the years 2010-2020 and annual wind speed in the years 2010, 2015, and 2020 separately, where $n$ is the sample size, and $95^{th}$ and $99^{th}$ refer to the $95^{th}$ and $99^{th}$ percentiles, respectively.  }\label{tab:table1_ds}
  \begin{subtable}{\linewidth}
    \centering
    \caption{Descriptive statistics of long-term wind speed at different heights}
    \begin{threeparttable}
\begin{tabular}{lllllllllll}
\hline
\begin{tabular}[c]{@{}l@{}}Height (m)\end{tabular} & 
\begin{tabular}[c]{@{}l@{}}$n$\end{tabular} &
\begin{tabular}[c]{@{}l@{}}Min\end{tabular} & 
\begin{tabular}[c]{@{}l@{}}Max\end{tabular} & 
\begin{tabular}[c]{@{}l@{}}Median\end{tabular} & 
\begin{tabular}[c]{@{}l@{}}Mean\end{tabular} & 
\begin{tabular}[c]{@{}l@{}}Variance\end{tabular} & 
\begin{tabular}[c]{@{}l@{}}Skewness\end{tabular} & 
\begin{tabular}[c]{@{}l@{}}Kurtosis\end{tabular} &
\begin{tabular}[c]{@{}l@{}}$95^{th}$\end{tabular} & 
\begin{tabular}[c]{@{}l@{}}$99^{th}$\end{tabular} \\
\hline
10  & 96,432 & 0.24 & 28.21 & 2.75  & 3.55 & 7.06 & 2.24 & 10.03 & 8.86 & 13.95\\ 
20  & 96,432 & 0.23 & 30.62 & 3.03  & 3.89 & 8.58 & 2.15 & 9.6 & 9.74 & 15.22\\ 
50  & 96,432 & 0.27 & 33.97 & 3.55  & 4.48 & 10.99 & 2.03 & 8.88 & 11.05 & 17.09\\ 
80  & 96,432 & 0.27 & 35.71 & 3.66  & 4.64 & 12.07 & 1.94 & 8.35 & 11.50 & 17.66\\ 
\hline
\end{tabular}
\end{threeparttable}
  \end{subtable}
 \vspace{0.3\baselineskip}

  \begin{subtable}{\linewidth}
  \caption{Descriptive statistics of annual wind speed at height 80 m }
    \centering
        \begin{threeparttable}
\begin{tabular}{lllllllllll}
\hline
\begin{tabular}[c]{@{}l@{}}Year\end{tabular} & 
\begin{tabular}[c]{@{}l@{}}$n$\end{tabular} &
\begin{tabular}[c]{@{}l@{}}Min\end{tabular} & 
\begin{tabular}[c]{@{}l@{}}Max\end{tabular} & 
\begin{tabular}[c]{@{}l@{}}Median\end{tabular} & 
\begin{tabular}[c]{@{}l@{}}Mean\end{tabular} & 
\begin{tabular}[c]{@{}l@{}}Variance\end{tabular} & 
\begin{tabular}[c]{@{}l@{}}Skewness\end{tabular} & 
\begin{tabular}[c]{@{}l@{}}Kurtosis\end{tabular} &
\begin{tabular}[c]{@{}l@{}}$95^{th}$\end{tabular} & 
\begin{tabular}[c]{@{}l@{}}$99^{th}$\end{tabular} \\
\hline
2010 & 8,760 & 0.27 & 26.7 & 3.32 & 4.21 & 10.34 & 2.03 & 8.56 & 10.74 & 16.59\\ 
2015 & 8,760 & 0.31 & 35.71 & 3.45 & 4.37 & 10.99 & 1.99 & 8.85 & 11.10 & 16.62 \\
2020 & 8,784 & 0.34 & 33.33 & 3.81 & 4.79 & 12.38 & 2.04 & 9.36 & 11.61 & 18.13 \\
\hline
\end{tabular}
\end{threeparttable}
  \end{subtable}
  % \begin{tablenotes}
  %       \footnotesize
  %       \item[*] $n$ is the sample size; $95^{th}$ and $99^{th}$ refer to the $95^{th}$ and $99^{th}$ percentiles, respectively. 
  %     \end{tablenotes}
\end{table*}
\end{landscape}

\begin{landscape}
  \begin{table*}
\centering
\caption{Notations and PDFs for BGL and other reference probability distributions}\label{tab:table1_pdf}
\begin{threeparttable}
\begin{tabular}{llll}
\hline
%\begin{tabular}[c]{@{}l@{}}\textbf{Wind Data Year}\end{tabular} & 
\begin{tabular}[c]{@{}l@{}}Distribution \end{tabular} & 
\begin{tabular}[c]{@{}l@{}}Notation\end{tabular} & 
\begin{tabular}[c]{@{}l@{}}$p$\end{tabular} & 
\begin{tabular}[c]{@{}l@{}}PDF\end{tabular} \\
\hline
Beta-generalized Lindley & BGL & 4 & $f_{BGL}(x; \alpha,\lambda, a, b) = \frac{\alpha \lambda^{2}(1+x)e^{-\lambda x}}{B(a,b)(1+\lambda)} \left[1-\frac{1+\lambda +\lambda x}{1+\lambda}e^{-\lambda x}\right]^{a\alpha -1} \left\{1-\left[1-\frac{1+\lambda +\lambda x}{1+\lambda}e^{-\lambda x}\right]^{\alpha}\right\}^{b-1}$  \\       %
Beta-Lindley & BL & 3 & $f_{BL}(x;\lambda, a, b)=\frac{\lambda^{2}(1+x)e^{-\lambda x}}{B(a,b)(1+\lambda)} \left[1-\frac{1+\lambda +\lambda x}{1+\lambda}e^{-\lambda x}\right]^{a -1} \left[\frac{1+\lambda +\lambda x}{1+\lambda}e^{-\lambda x}\right]^{b-1}$ \\  
Generalized Lindley & GL & 2 & $f_{GL}(x;\alpha,\lambda)=\frac{\alpha \lambda^{2}}{1+\lambda}(1+x)\left[1-\frac{1+\lambda +\lambda x}{1+\lambda}e^{-\lambda x}\right]^{\alpha-1}e^{-\lambda x}$  \\  
Lindley & L & 1 & $f_{L}(x;\lambda)=\frac{\lambda^{2}}{1+\lambda}(1+x)e^{-\lambda x}$  \\  
Gamma & GAM & 2 & $f_{GAM}(x;\alpha,\lambda) = \frac{\lambda^\alpha}{\Gamma(\alpha)}x^{\alpha-1}e^{-\lambda x}$    \\  
Beta-Weibull & BW & 4  & $f_{BW}(x; \alpha, \lambda, a, b)=\frac{\alpha \lambda^{\alpha}}{B(a,b)}x^{\alpha-1}e^{-b(\lambda x)^{\alpha}}[1-e^{-(\lambda x)^{\alpha}}]^{a-1}$ \\  
Weibull & W & 2 & $f_{W}(x; \alpha, \lambda)=\alpha \lambda^{\alpha}x^{\alpha-1}e^{-(\lambda x)^{\alpha}}$  \\  
%Rayleigh & $R$ & 1 & $f_{R}(x; \lambda)=2 \lambda^{2}x e^{-(\lambda x)^{2}}$ \\
Beta-Exponential & BE  & 3 & $f_{BE}(x; \lambda,a, b)= \frac{\lambda}{B(a,b)} e^{-b\lambda x}(1-e^{-\lambda x})^{a-1}$ \\  
Log-Normal & LogN & 2 & $f_{LogN}(x;\alpha,\lambda) = \frac{1}{x \lambda \sqrt{2\pi}} e^{-\frac{(ln x - \alpha)^2}{2\lambda^2}}$  \\  
\hline
\end{tabular}
\begin{tablenotes}
        \footnotesize
        \item[*] $p$ is the number of unknown parameters for each distribution, and PDF refers to the probability density function. 
      \end{tablenotes}
\end{threeparttable}
\end{table*}
\end{landscape}

\begin{landscape}
 \begin{table*}%[width=15.5cm]
 \small
\centering
\caption{The parameter estimates, $-2\ln(L)$, AIC, BIC, KS statistic, and AD statistic of BGL distribution and other reference models for wind speed data in the years 2010-2020 at heights 10, 20, 50, and 80 m.}\label{tab:table1_H10}
\begin{threeparttable}
\begin{tabular}{lllllllllll}
\hline
\begin{tabular}[c]{@{}l@{}}Hieght (m)\end{tabular} & 
\begin{tabular}[c]{@{}l@{}}Distribution \end{tabular} & 
\begin{tabular}[c]{@{}l@{}}$\alpha$ \end{tabular} & 
\begin{tabular}[c]{@{}l@{}}$\lambda$ \end{tabular} & 
\begin{tabular}[c]{@{}l@{}}$a$ \end{tabular} & 
\begin{tabular}[c]{@{}l@{}}$b$ \end{tabular} & 
\begin{tabular}[c]{@{}l@{}}$-2\ln(L)$ \end{tabular} & 
\begin{tabular}[c]{@{}l@{}}AIC \end{tabular} & 
\begin{tabular}[c]{@{}l@{}}BIC \end{tabular} & 
\begin{tabular}[c]{@{}l@{}}KS\end{tabular} &
\begin{tabular}[c]{@{}l@{}}AD \end{tabular} 
\\
\hline
\multirow{9}{3em}{10} & $BGL(\alpha,\lambda, a, b)$ & 46.822 & 1.063 & 0.081 & 0.349 & \textbf{393606.3} & \textbf{393614.3} &  \textbf{393652.2} & \textbf{0.008} &  \textbf{11.4}  \\   
&$LogN(\alpha,\lambda)$ & 1.047 & 0.659 & - & - & 395230.1 & 395234.1 & 395253.0 & 0.026 & 125.6  \\ 
&$BW(\alpha,\lambda, a, b)$ & 0.156 &  0.706 & 105.217 & 51.390 & 395454.2 & 395462.2 & 395500.1  & 0.031 & 172.8 \\  
&$BE(\lambda,a, b)$  & - & 1.657 & 5.888 & 0.254 & 395777.9 & 395783.9 & 395812.4 & 0.036 & 206.3 \\  
&$BL(\lambda, a, b)$ & - & 2.258 & 5.450 & 0.202 & 396153.3 &  396159.3 &  396187.8  & 0.040 & 260.1 \\  
&$GAM(\alpha,\lambda)$ & 2.423 & 0.682  & - & - & 403413.3 &  403417.3 & 403436.2 & 0.073 & 953.2 \\  
&$GL(\alpha,\lambda)$ & 2.038 & 0.663  & - & - & 403861.7 & 403865.7 & 403884.6 & 0.073 & 985.2  \\  
&$W(\alpha,\lambda)$ & 1.500 & 0.252 & - & - & 412722.1 & 412726.1 & 412745.1 & 0.083 & 1733 \\  
&$L(\lambda)$ & - & 0.473 & - & - & 419689.8 & 419691.8 & 419701.3 & 0.136 & 3079.3 \\  
\hline

\multirow{9}{3em}{20} & $BGL(\alpha,\lambda, a, b)$ & 41.012 & 0.803 & 0.064 & 0.417 & \textbf{420042.8} & \textbf{420050.8} & \textbf{420088.7} & \textbf{0.019} &  \textbf{73.9}   \\       %($n$\tnote{*}=96432) 
&$BW(\alpha,\lambda, a, b)$ & 0.405 & 0.153 & 13.107 & 11.898 & 422274.7 & 422282.7 & 422320.6 & 0.032 & 249.5  \\  
&$BE(\lambda,a, b)$  & - & 0.996 & 3.429 & 0.386 & 422286.2 & 422292.2 & 422320.7 & 0.038 & 271.7 \\  
&$BL(\lambda, a, b)$ & - & 1.526 & 3.014 & 0.277 & 422446.6 & 422452.6 & 422481.0 & 0.042 &  299.1 \\  
&$LogN(\alpha,\lambda)$ & 1.119 & 0.711 & - & - & 423794.1 & 423798.1 & 423817.1 & 0.035 &  258.1 \\  
&$GAM(\alpha,\lambda)$ & 2.242 & 0.576 & - & - & 425813.2 & 425817.2 & 425836.2 & 0.062 & 668.8   \\  
&$GL(\alpha,\lambda)$ & 1.785 & 0.576 & - & - & 426342.3 & 426346.3 & 426365.2 & 0.065 & 708.2  \\  
&$W(\alpha,\lambda)$ & 1.473 & 0.231 & - & - & 432651.2 & 432655.2 & 432674.2 & 0.075 & 1323.4 \\  
&$L(\lambda)$ & - & 0.436 & - & - & 437665.8 & 437667.8 & 437677.3 & 0.119 & 2387.3 \\  
\hline

\multirow{9}{3em}{50} & $BGL(\alpha,\lambda, a, b)$ & 33.374 &  0.704 & 0.077 & 0.438 & \textbf{447630.4} & \textbf{447638.4}  & \textbf{447676.3} & \textbf{0.007} &  \textbf{13.6}   \\       %($n$\tnote{*}=96432) 
&$BW(\alpha,\lambda, a, b)$ & 0.393 &  0.155 & 14.214 & 11.664 & 448928.0 & 448936.0 & 448973.9 & 0.023 &  103.5 \\  
&$BE(\lambda,a, b)$  & - & 0.918 & 3.599 & 0.361 & 448946.1 & 448952.1 & 448980.6 & 0.027 & 119.4 \\  
&$BL(\lambda, a, b)$ & - & 1.425 & 3.146 & 0.255 &  449099.9 & 449105.9 & 449134.3 & 0.031 & 140 \\  
&$LogN(\alpha,\lambda)$ & 1.264 & 0.706 & - & - & 450128.5 & 450132.5 & 450151.5 & 0.025 & 114.4 \\  
&$GAM(\alpha,\lambda)$ & 2.270 & 0.507 & - & - & 452344.0 & 452348.0 & 452366.9 & 0.053 &  475.7 \\  
&$GL(\alpha,\lambda)$ & 1.749 & 0.503 & - & - & 453028.5 & 453032.5 & 453051.5 & 0.056 & 532 \\  
&$W(\alpha,\lambda)$ & 1.490 & 0.200  & - & - & 458927.6 & 458931.6 & 458950.5 & 0.069 & 1067.3 \\  
&$L(\lambda)$ & - & 0.384 & - & - & 463639.8 & 463641.8 & 463651.3 & 0.105 &  2037.5 \\  
\hline
\multirow{9}{3em}{80}&$BGL(\alpha,\lambda, a, b)$ & 25.281 & 0.669 & 0.093 & 0.456 & \textbf{458385.4} & \textbf{458393.4} & \textbf{458431.4} & \textbf{0.006} &  \textbf{7.6}  \\       %($n$\tnote{*}=96432) 
&$BE(\lambda,a, b)$  & - & 0.882 & 3.279 & 0.353 & 459224.1 & 459230.1 & 459258.6 & 0.022 & 75.0 \\  
&$BW(\alpha,\lambda, a, b)$ & 0.463 & 0.175 & 9.712 & 7.680 & 459243 & 459251 & 459289 & 0.019 & 66.9 \\  
&$BL(\lambda, a, b)$ & - & 1.415 & 2.894 & 0.241 & 459350.2 & 459356.2 & 459384.6 & 0.026 & 90.4  \\  
&$LogN(\alpha,\lambda)$ & 1.285 & 0.730 & - & - & 460878.1 & 460882.1 & 460901.1 & 0.024 & 108.7 \\  
&$GAM(\alpha,\lambda)$ & 2.158 & 0.465 & - & - & 462015.0 & 462019.0 & 462037.9 & 0.047 &  363.8 \\  
&$GL(\alpha,\lambda)$ & 1.625 & 0.472  & - & - & 462985.3 & 462989.3 & 463008.3 & 0.052 & 449.5 \\  
&$W(\alpha,\lambda)$ & 1.463 & 0.194 & - & - & 467606.4 & 467610.4 & 467629.3 & 0.062 & 852.2 \\  
&$L(\lambda)$ & - & 0.373 & - & - & 471292.1 & 471294.1 & 471303.6 & 0.091 & 1574.9 \\  
\hline
\end{tabular}
\begin{tablenotes}
        \footnotesize
        \item[*] The distribution at each height is listed in the ascending order of $-2\ln(L)$. The best results are marked in bold. Refer to Table \ref{tab:table1_pdf} for the notation and PDF of each distribution.  
      \end{tablenotes}
\end{threeparttable}
\end{table*}

\end{landscape}

\begin{table*}
\centering
\caption{The observed (obs) and estimated (est) $95^{th}$ and $99^{th}$ percentiles for wind speed for the years 2010-2020  at heights 10, 20, 50, and 80 m}\label{tab:table1_percentile}
\begin{threeparttable}
\begin{tabular}{llllllllll}
\hline
\begin{tabular}[c]{@{}l@{}}Height (m)\end{tabular} & 
\begin{tabular}[c]{@{}l@{}}Distribution\end{tabular} & 
\begin{tabular}[c]{@{}l@{}}$95^{th}$ obs\end{tabular} & 
\begin{tabular}[c]{@{}l@{}}$95^{th}$ est\end{tabular} & 
\begin{tabular}[c]{@{}l@{}}Bias\end{tabular} & 
\begin{tabular}[c]{@{}l@{}}$99^{th}$ obs\end{tabular} & 
\begin{tabular}[c]{@{}l@{}}$99^{th}$ est\end{tabular} & 
\begin{tabular}[c]{@{}l@{}}Bias\end{tabular}  \\
\hline
\multirow{9}{4em}{10} & $BGL(\alpha,\lambda, a, b)$ & 8.86 & 8.93 & \textbf{0.07} & 13.95 & 13.60 & \textbf{-0.35} \\   
&$L(\lambda)$ & 8.86 & 9.25 & 0.39 & 13.95 & 13.25 & -0.70 \\  
&$LogN(\alpha,\lambda)$ & 8.86 & 8.43 & -0.43 & 13.95 & 13.20 & -0.75  \\ 
&$BE(\lambda,a, b)$  &  8.86 & 8.38 & -0.48 & 13.95 & 12.19 & -1.76 \\
&$BL(\lambda, a, b)$ &  8.86 & 8.33 & -0.53 & 13.95 & 12.01 & -1.94 \\  
&$BW(\alpha,\lambda, a, b)$ &  8.86 & 8.32 & -0.54 & 13.95 & 12.75 & -1.20 \\
&$W(\alpha,\lambda)$ & 8.86 & 8.25 & -0.61 & 13.95 & 10.99 & -2.96 \\   
&$GAM(\alpha,\lambda)$ &  8.86 & 7.94 & -0.92 & 13.95 & 10.86 & -3.09 \\
&$GL(\alpha,\lambda)$ &   8.86 & 7.69 & -1.17 & 13.95 & 10.50 & -3.45 \\
\hline
\multirow{9}{4em}{20} & $BGL(\alpha,\lambda, a, b)$ & 9.74 & 9.77 & \textbf{0.03} & 15.22 & 15.01 & \textbf{-0.21} \\
&$LogN(\alpha,\lambda)$ & 9.74 & 9.87 & 0.13 & 15.22 & 16.01 & 0.79 \\ 
&$L(\lambda)$ &  9.74 & 10.08 & 0.34 & 15.22 & 14.42 & -0.80 \\
&$BW(\alpha,\lambda, a, b)$ &  9.74 & 9.31 & -0.43 & 15.22 & 14.00 & -1.22 \\
&$BE(\lambda,a, b)$  &  9.74 & 9.26 & -0.48 & 15.22 & 13.46 & -1.76 \\
&$BL(\lambda, a, b)$ &  9.74 & 9.20 & -0.54 & 15.22 & 13.21 & -2.01 \\  
&$W(\alpha,\lambda)$ &  9.74 & 9.13 & -0.61 & 15.22 & 12.22 & -3.00 \\
&$GAM(\alpha,\lambda)$ &  9.74 & 8.90 & -0.84 & 15.22 & 12.29 & -2.93 \\   
&$GL(\alpha,\lambda)$ &  9.74 & 8.66 & -1.08 & 15.22 & 11.90 & -3.32 \\ 
\hline
\multirow{9}{4em}{50} & $BGL(\alpha,\lambda, a, b)$ &  11.05 & 11.13 & \textbf{0.08} & 17.09 & 16.83 & \textbf{-0.26} \\  
&$LogN(\alpha,\lambda)$ &  11.05 & 11.29 & 0.24 & 17.09 & 18.27 & 1.18 \\
&$BW(\alpha,\lambda, a, b)$ &  11.05 & 10.71 & -0.34 & 17.09 & 16.12 & -0.97 \\
&$BE(\lambda,a, b)$  &  11.05 & 10.68 & -0.37 & 17.09 & 15.53 & -1.56 \\
&$BL(\lambda, a, b)$ &  11.05 & 10.62 & -0.43 & 17.09 & 15.27 & -1.82 \\
&$L(\lambda)$ &  11.05 & 11.53 & 0.48 & 17.09 & 16.45 & -0.64 \\ 
&$W(\alpha,\lambda)$ & 11.05 & 10.44 & -0.61 & 17.09 & 13.93 & -3.16 \\
&$GAM(\alpha,\lambda)$ &  11.05 & 10.22 & -0.83 & 17.09 & 14.08 & -3.01 \\
&$GL(\alpha,\lambda)$ &  11.05 & 9.96 & -1.09 & 17.09 & 13.68 & -3.41 \\
\hline
\multirow{9}{4em}{80} & $BGL(\alpha,\lambda, a, b)$ & 11.50 & 11.60 & \textbf{0.10} & 17.66 & 17.38 & \textbf{-0.28} \\ 
&$BW(\alpha,\lambda, a, b)$ &  11.50 & 11.25 & -0.25 & 17.66 & 16.95 & -0.71 \\
&$BE(\lambda,a, b)$  &  11.50 & 11.23 & -0.27 & 17.66 & 16.40 & -1.26 \\
&$BL(\lambda, a, b)$ &  11.50 & 11.16 & -0.34 & 17.66 & 16.12 & -1.54 \\
&$L(\lambda)$ &  11.50 & 11.91 & 0.41 & 17.66 & 16.98 & -0.68 \\
&$LogN(\alpha,\lambda)$ &  11.50 & 12.02 & 0.52 & 17.66 & 19.77 & 2.11 \\
&$W(\alpha,\lambda)$ &  11.50 & 10.92 & -0.58 & 17.66 & 14.65 & -3.01 \\
&$GAM(\alpha,\lambda)$ &  11.50 & 10.74 & -0.76 & 17.66 & 14.89 & -2.77 \\
&$GL(\alpha,\lambda)$ &  11.50 & 10.48 & -1.02 & 17.66 & 14.45 & -3.21\\
\hline
\end{tabular}
\begin{tablenotes}
        \footnotesize
        \item[*] Bias = estimated percentile - observed percentile. The estimated percentile is from the fitted distribution, and the observed percentile is from the original wind speed data. The distribution at each height is listed in the ascending order of absolute value of bias for the $95^{th}$ percentile score. The best results are marked in bold. Refer to Table \ref{tab:table1_pdf} for the notation and PDF of each distribution.
      \end{tablenotes}
\end{threeparttable}
\end{table*}

\begin{landscape}
\begin{table*}%[width=14 cm]
\centering
\caption{The parameter estimates, $-2\ln(L)$, AIC, BIC, KS statistic, and AD statistic of BGL distribution and other reference models for wind speed data in the years 2010, 2015 and 2020 at height 80 m.}\label{tab:table_H80}
\begin{threeparttable}
\begin{tabular}{lllllllllll}
\hline
\begin{tabular}[c]{@{}l@{}}Year\end{tabular} & 
\begin{tabular}[c]{@{}l@{}}Distribution \end{tabular} & 
\begin{tabular}[c]{@{}l@{}}$\alpha$ 
\end{tabular} & 
\begin{tabular}[c]{@{}l@{}}$\lambda$ 
\end{tabular} & 
\begin{tabular}[c]{@{}l@{}}$a$ 
\end{tabular} & 
\begin{tabular}[c]{@{}l@{}}$b$ 
\end{tabular} & 
\begin{tabular}[c]{@{}l@{}}$-2\ln(L)$ \end{tabular} & 
\begin{tabular}[c]{@{}l@{}}AIC  
\end{tabular} & 
\begin{tabular}[c]{@{}l@{}}BIC 
\end{tabular} &
\begin{tabular}[c]{@{}l@{}}KS 
\end{tabular} & 
\begin{tabular}[c]{@{}l@{}}AD 
\end{tabular} 
\\
\hline
%Year 2010  
\multirow{9}{2em}{2010} & $BGL(\alpha,\lambda, a, b)$ &  32.951 &  0.730 & 0.073 & 0.429 & \textbf{39909.7} & \textbf{39917.7} & \textbf{39946} & \textbf{0.005} & \textbf{0.25} \\  
&$BW(\alpha,\lambda, a, b)$ &  0.364  & 0.157 & 15.394 & 12.811 & 39989.1 & 39997.1 & 40025.4 & 0.021 & 5.33 \\
&$LogN(\alpha,\lambda)$ & 1.184 & 0.723  & - & - &  40085.1 & 40089.1 & 40103.3 & 0.024 & 7.91 \\  
&$BE(\lambda,a, b)$  & - &  0.0450 & 2.132 & 9.620 & 40317.8 & 40323.8 & 40345 & 0.051 & 38.28 \\
&$GAM(\alpha,\lambda)$ & 2.132 & 0.507 & - & - &  40321.9 & 40325.9 & 40340.1 & 0.052 & 38.58 \\  
&$GL(\alpha,\lambda)$ &  1.642 & 0.518    & - & - &  40423.2 & 40427.2 & 40441.4 & 0.055 & 46.46 \\
&$W(\alpha,\lambda)$ & 1.444 & 0.214  & - & - &  40870.8 & 40874.8 & 40888.9 & 0.067 & 86.91 \\
&$L(\lambda)$ & - & 0.407 & - & - &   41206 & 41208 & 41215.1 & 0.094 & 153.26 \\
&$BL(\lambda, a, b)$ & - &  0.056 & 1.018 & 24.390  & 41318.9 & 41324.9 & 41346.2 & 0.089 & 126.97 \\ 
\hline
\multirow{9}{2em}{2015} & $BGL(\alpha,\lambda, a, b)$ & 29.330 & 0.667  & 0.075 & 0.467 & \textbf{40775.6} & \textbf{40783.6} & \textbf{40811.9} & \textbf{0.010} & \textbf{1.83} \\
&$BW(\alpha,\lambda, a, b)$ & 0.493 & 0.079 & 8.039 & 11.294 & 40876.4 & 40884.4 & 40912.7 & 0.023 & 9.56 \\  
&$BL(\lambda, a, b)$ & - & 1.385 & 2.591 & 0.261 & 40887.5 & 40893.5 & 40914.7 & 0.029 & 11.55 \\  
&$BE(\lambda,a, b)$  & - & 0.054 & 2.101 & 8.371 & 41081.8 & 41087.8 & 41109.1 & 0.048 & 31.98 \\ 
&$LogN(\alpha,\lambda)$ & 1.217 & 0.748  & - & - &  41086.1 & 41090.1 & 41104.3 & 0.034 & 18.39 \\ 
&$GAM(\alpha,\lambda)$ & 2.093 & 0.479  & - & - &  41086.1 & 41090.1 & 41104.3 & 0.048 & 32.37 \\ 
&$GL(\alpha,\lambda)$ & 1.586 & 0.492  & - & - & 41171 & 41175 & 41189.2 & 0.053 & 39.31 \\ 
&$W(\alpha,\lambda)$ &  1.444 & 0.206  & - & - &  41554.9 & 41558.9 & 41573 & 0.063 & 75.39 \\ 
&$L(\lambda)$ & - & 0.393 & - & - & 41863.7 & 41865.7 & 41872.8 & 0.090 & 138.20 \\ 
\hline
\multirow{9}{2em}{2020} & $BGL(\alpha,\lambda, a, b)$ & 29.649 & 0.636 & 0.081 & 0.472 & \textbf{42058.5} & \textbf{42066.5} & \textbf{42094.8} & \textbf{0.009} & \textbf{1.31} \\  
&$BE(\lambda,a, b)$  & - & 0.782 & 3.338 & 0.399 & 42146.6 & 42152.6 & 42173.9 & 0.023 & 8.22 \\ 
&$BW(\alpha,\lambda, a, b)$ &  0.444 &  0.058 & 10.729 & 15.713 & 42148.2 & 42156.2 & 42184.5 & 0.023 & 7.78 \\  
&$BL(\lambda, a, b)$ & - & 1.269 & 2.885 & 0.270 & 42155.6 & 42161.6 & 42182.9 & 0.028 & 9.31 \\  
&$LogN(\alpha,\lambda)$ & 1.328 & 0.713 & - & - &  42306.3 & 42310.3 & 42324.4 & 0.027 & 12.44 \\
&$GAM(\alpha,\lambda)$ &  2.257 & 0.472 & - & - &  42390.9 & 42394.9 & 42409.1 & 0.049 & 32.98 \\ 
&$GL(\alpha,\lambda)$ & 1.706 & 0.469 & - & - & 42449.2 & 42453.2 & 42467.4 & 0.051 & 37.72 \\ 
&$W(\alpha,\lambda)$ & 1.494 & 0.187  & - & - &  42945.5 & 42949.5 & 42963.7 & 0.065 & 82.09 \\
&$L(\lambda)$ & - & 0.362 & - & - &  43344.9 & 43346.9 & 43354 & 0.101 & 167.16 \\
\hline 
\end{tabular}
\begin{tablenotes}
        \footnotesize
        \item[*] The distribution for each year is listed in the ascending order of $-2\ln(L)$. The best results are marked in bold. Refer to Table \ref{tab:table1_pdf} for the notation and PDF of each distribution.
      \end{tablenotes}
\end{threeparttable}
\end{table*}
\end{landscape}

\section*{Funding}
This research receives no grant from any funding agency in the public, commercial, or not-for-profit sectors.

\section*{Acknowledgments}
The authors thank Dr. Senfeng Liu for effective discussions. We provide the code of BGL distribution and the preprocessed wind speed data at the GitHub page \url{https://github.com/tiantiy/BGL-for-Wind}

\section*{Competing interest}
The authors declare no competing interests for this paper.

\section*{Authorship Contribution}
Tiantian Yang: methodology, algorithm implementation, data analysis, formal analysis, visualization, manuscript preparation, manuscript submission. 

Dongwei Chen: conceptualization, data acquisition, application design, formal analysis, methodology, manuscript preparation, manuscript submission.

\bibliographystyle{plain} 
\bibliography{refs}

\end{document}